# Velocity Saturation in La-doped $BaSnO_3$ Thin Films

Hareesh Chandrasekar,[1,a)] Junao Cheng,[1] Tianshi Wang,[3] Zhanbo Xia,[1] Nicholas G. Combs,[2] Christopher R. Freeze,[2] Patrick B. Marshall,[2] Joe McGlone,[1] Aaron Arehart,[1] Steven Ringel,[1,4] Anderson Janotti,[3] Susanne Stemmer,[2] Wu Lu[1] and Siddharth Rajan[1,4]

[1] Department of Electrical and Computer Engineering, The Ohio State University, Columbus, OH 43210, USA

[2] Materials Department, University of California, Santa Barbara, CA 93106, USA

[3] Department of Materials Science and Engineering, University of Delaware, Newark, DE 19716, USA

[4] Department of Material Science Engineering, The Ohio State University, Columbus, OH 43210, USA

$BaSnO_3$, a high mobility perovskite oxide, is an attractive material for oxide-based electronic devices. However, in addition to low-field mobility, high-field transport properties such as the saturation velocity of carriers play a major role in determining device performance. We report on the experimental measurement of electron saturation velocity in La-doped $BaSnO_3$ thin films for a range of doping densities. Predicted saturation velocities based on a simple LO-phonon emission model using an effective LO phonon energy of 120 meV show good agreement with measurements of velocity saturation in La-doped $BaSnO_3$ films. Density-dependent saturation velocity in the range of $1.6\times10^7$ cm/s reducing to $2\times10^6$ cm/s is predicted for δ-doped $BaSnO_3$ channels with carrier densities ranging from $10^{13}$ cm$^{-2}$ to $2\times10^{14}$ cm$^{-2}$ respectively. These results are expected to aid the informed design of $BaSnO_3$ as the active material for high-charge density electronic transistors.

a) Electronic Mail: chandrasekar.28@osu.edu

Perovskite oxides continue to attract a lot of interest for fundamental and applied physics in view of the plethora of electronic properties they exhibit.[1, 2] This also makes perovskites an ideal platform to integrate diverse device functionalities onto. However, the poor electron mobility of most perovskite oxides makes them a sub-optimal choice for electronic device applications. With a recorded room-temperature mobility of 320 $cm^2/Vs$ in single-crystals,[3] $BaSnO_3$ (BSO) is a notable exception to this rule. The cause for such high mobility in BSO thin films has been studied previously and attributed both to the low effective mass of electrons owing to the significant anti-bonding *s*- orbital-like character of the conduction-band minimum,[4] and the low electron-phonon scattering rate due to the lower density of conduction band states.[5] Such high mobilities in conjunction with the large and tunable doping densities ($\sim 10^{20}$ $cm^{-3}$) achievable in this material, and the possibility of realizing hetero-structures with carrier confinement makes BSO promising from an electronic device perspective.[6] Transistors fabricated from La-doped $BaSnO_3$ and $SrSnO_3$ channels (10-28 nm) are still in the early stages of development but have shown promising results compared to those based on other perovskite-oxides such as $SrTiO_3$.[7-9] However, in addition to the low-field transport properties such as mobility, the operation of electronic devices such as RF amplifiers and highly-scaled transistors for logic depends critically on high-field transport, i.e. saturation velocity of charge carriers in the high-mobility channel.[10,11] While low-field transport in BSO has been studied and modeled, there are no reports on high-field transport and the physical mechanisms limiting saturation velocity in this material, despite its relative importance for device performance. Here we report on the experimental measurement of saturation velocity in thin films of La-doped BSO and model its dependence on the dominant LO-phonon scattering process. We obtain reasonable agreement between the theoretical values based on a simple optical phonon emission model and measured saturation velocities, for an experimentally relevant range of doping densities in BSO thin films.

**TABLE I.** Sample stack and thicknesses of the MBE-grown La-doped BSO thin films used in this study along with their measured sheet carrier concentrations ($cm^{-2}$) and Hall mobilities ($cm^2/Vs$).

| Sample Details | Sheet Carrier Concentration ($cm^{-2}$) | Hall Mobility ($cm^2/Vs$) |
|---|---|---|
| **BTO/La:BSO/BSO/DSO (20 nm/21 nm/11 nm/substrate)** | $2.65\times10^{14}$ | 96 |
| **BTO/La:BSO/BSO/STO (20 nm/16 nm/5 nm/substrate)** | $1.92\times10^{14}$ | 87 |
| **La:BSO/BSO/STO (20 nm/5 nm/substrate)** | $7.23\times10^{13}$ | 58 |

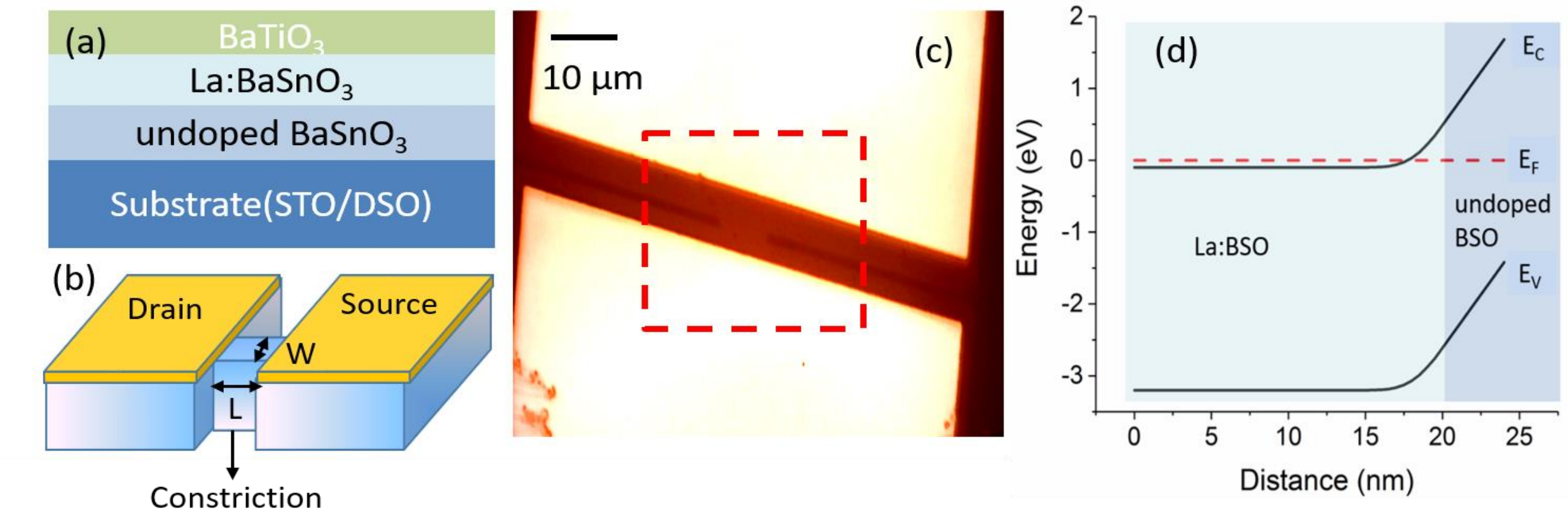

**FIG. 1.** (a) Cross-section of a representative La:BSO thin film stack used in this study, (b) Schematic of the I-shaped test structure used for saturation velocity measurements with width 'W' and length 'L' as indicated, (c) Optical image of a fabricated test structure with the constriction mesa visible within the red dotted square (L=0.9 μm, W=10 μm). (d) Band diagram of 20 nm thick $10^{20}$ $cm^{-3}$ doped La-doped BSO – 5 nm thick un-doped BSO stacks (Fermi level at mid-gap), similar to those used in this study showing electron confinement almost throughout the entire La:BSO layer (depletion width <2.5 nm).

BSO films were grown via MBE with various La concentrations on insulating (100) $SrTiO_3$ (STO) and (110) $DyScO_3$ (DSO) substrates, [12] with and without $BaTiO_3$ (BTO) capping layers, as summarized in Table I. Due to the lattice mismatch between BSO and STO/DSO, all films are relaxed. For samples with a BTO capping layer, 20 nm of BTO was grown on the La-doped BSO film via hybrid MBE.[13] Saturation velocity was measured on specially designed I-shaped test structures consisting of wide source and drain contact pads with a narrow mesa-defined channel of length 'L' and width 'W' (see Fig. 1(b)) connecting them. Such a structure not only confines the current flow and applied potential to the channel, but also reduces any extraneous voltage drop in the contact/access regions due to the large difference in areas between the channel and contact regions.[14] Device fabrication consisted of i-line stepper lithography for all layers commencing with Ti/Al alignment markers, inductively coupled plasma-reactive ion etching (ICP-RIE) of the BTO cap layer using $BCl_3$/Ar, followed by ohmic contact deposition (Ti/Au, 50 nm/100 nm) using e-beam evaporation and finishing with an ICP-RIE ($BCl_3$/Ar) mesa-etch of the BSO channel to define the I-shaped constrictions. Various constriction widths (2 μm, 5 μm and 10 μm) and lengths (600 nm to 2 μm) were fabricated and tested on each sample. Pulsed I-V measurements (1 μs pulse width, 0.1% duty cycle) were performed using a Keithley 4200 semiconductor characterization system on the test structures. Contact resistances of 0.34 Ω.mm, 0.49 Ω.mm and 0.3 Ω.mm were measured on the three samples listed in Table I respectively using TLM test structures. Saturation velocity was extracted as $v_{sat} = J_{max} /qn_s$, where $J_{max}$ and $n_s$ are measured maximum saturation current density (A/mm) and sheet carrier concentration ($cm^{-2}$) respectively, and plotted as a function of electric field as shown in Fig. 2 for the sample with $n_s$=$2.65\times10^{14}$ $cm^{-2}$ as an example (Figs. S2 of the supplementary material show electric field vs current density plots for the complete set of measured devices across all samples used in this study). Saturation current densities ranging from 5-50 A/mm were observed for the range of sample doping densities. We also observed device-device variation across each sample comparable to that shown in Fig. 2, which we attribute to local doping/stoichiometric variations and structural defects in the deposited films. The critical thickness for strain relaxation for BSO films on STO and DSO substrates have been reported to be ~1 nm leading to structural defects in these films (threading dislocation densities $>10^{11}$ cm$^{-2}$).[15,16] In addition, pulsed I-V measurements were also performed for a subset of devices at 80 K. A comparison between electric field vs current density plots at 300 K and 80 K for two of the measured devices on the highest doping density sample is shown in Fig. 2 (b). An increase in $J_{max}$ (and hence extracted $v_{sat}$) of <18% can be observed for these two devices, which is comparable to the standard error of the mean (12%) measured across all devices at 300 K indicating that the current saturation is reflective of velocity and not mobility-limited. Fig S3 of supplementary material also shows electric field vs current density plot for the lowest doping density sample used in this study – $n_s = 7.23\times10^{13}$ cm$^{-2}$ showing <6% increase in $J_{max}$ and $v_{sat}$ at 80 K as compared to 300 K.

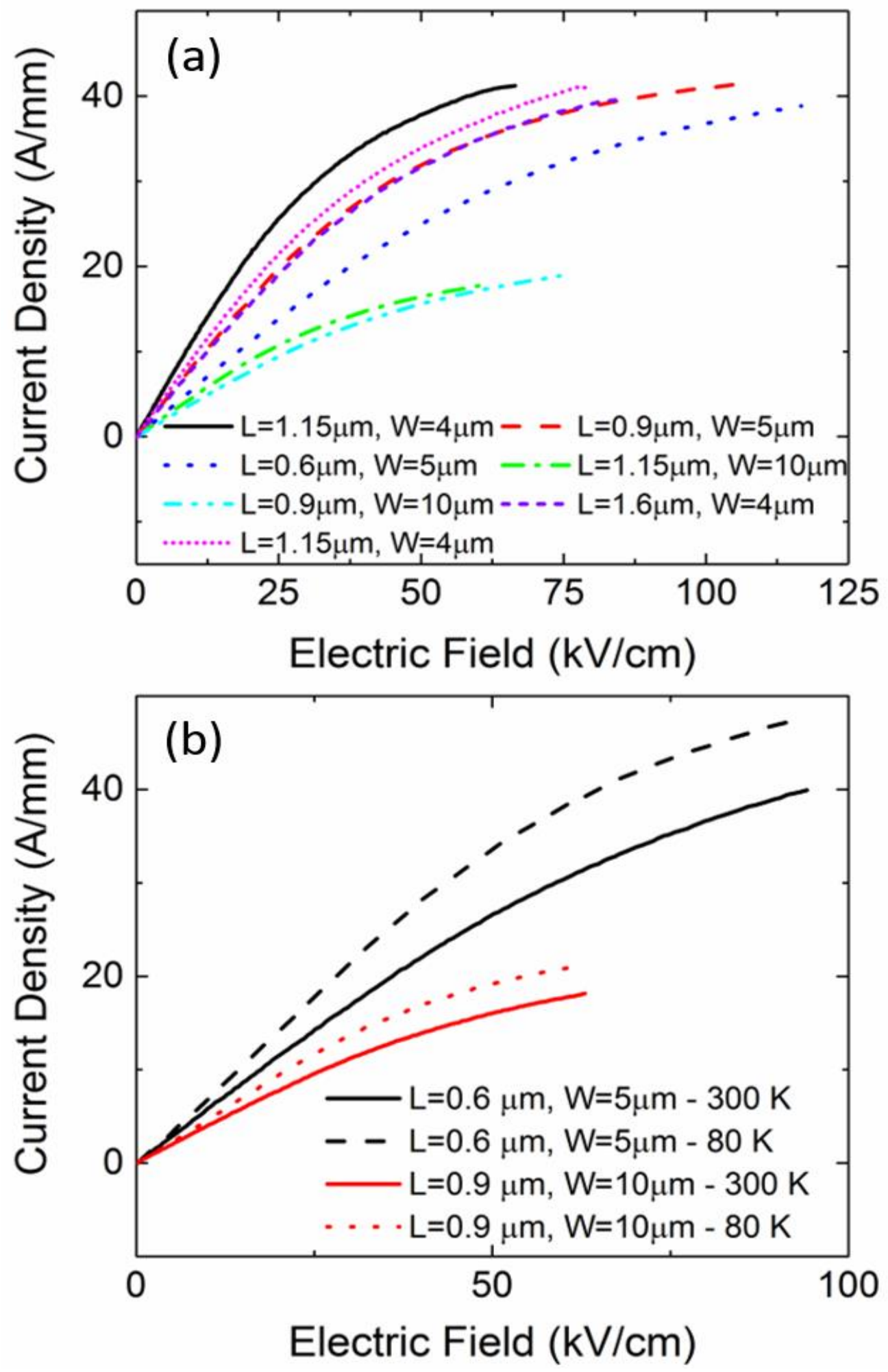

**FIG. 2.** Current density (A/mm) vs electric field (kV/cm) curves for a representative La:BSO film with doping density of $2.65\times10^{14}$ cm$^{-2}$ ($n_{3D}$ = $1.33\times10^{20}$ cm$^{-3}$) for the test structure dimensions indicated as measured at (a) 300 K across 7 devices, and (b) a comparison of 300 K and 80 K measured data for two devices on this sample.

In order to explain the mechanism of velocity saturation in these films, it is important to note that strong LO-phonon induced carrier scattering occurs in such polar materials.[5, 17] We performed first-principles calculations for the phonon spectrum and electron-phonon scattering rates in BSO based on density functional theory (DFT)[18,19] and density functional perturbation theory (DFPT)[20] as implemented in the Quantum ESPRESSO code,[21] with the generalized gradient approximation of Perdew-Burke-Ernzerhof for the exchange-correlation term.[22] Optimized norm-conserving Vanderbilt pseudopotentials were employed.[23] The EPW code was used to interpolate the electron-phonon matrix elements and calculate the electron-phonon scattering rates.[24] The phonon spectra of BSO (Fig.S3 in the supplementary material) exhibits 15 branches with the three LO phonon branches (branches 6, 12 and 15) strongly affecting electron-phonon scattering for long-range Frölich interactions. This can be clearly seen in Fig. 3 which plots the scattering rate for the various phonon branches as a function of electron energy with reference to the Fermi level. It is evident that the total scattering rate rises steeply beyond electron energies of 120 meV which may be taken as the net effective LO phonon energy for electron scattering due to optical phonon emission in BSO.

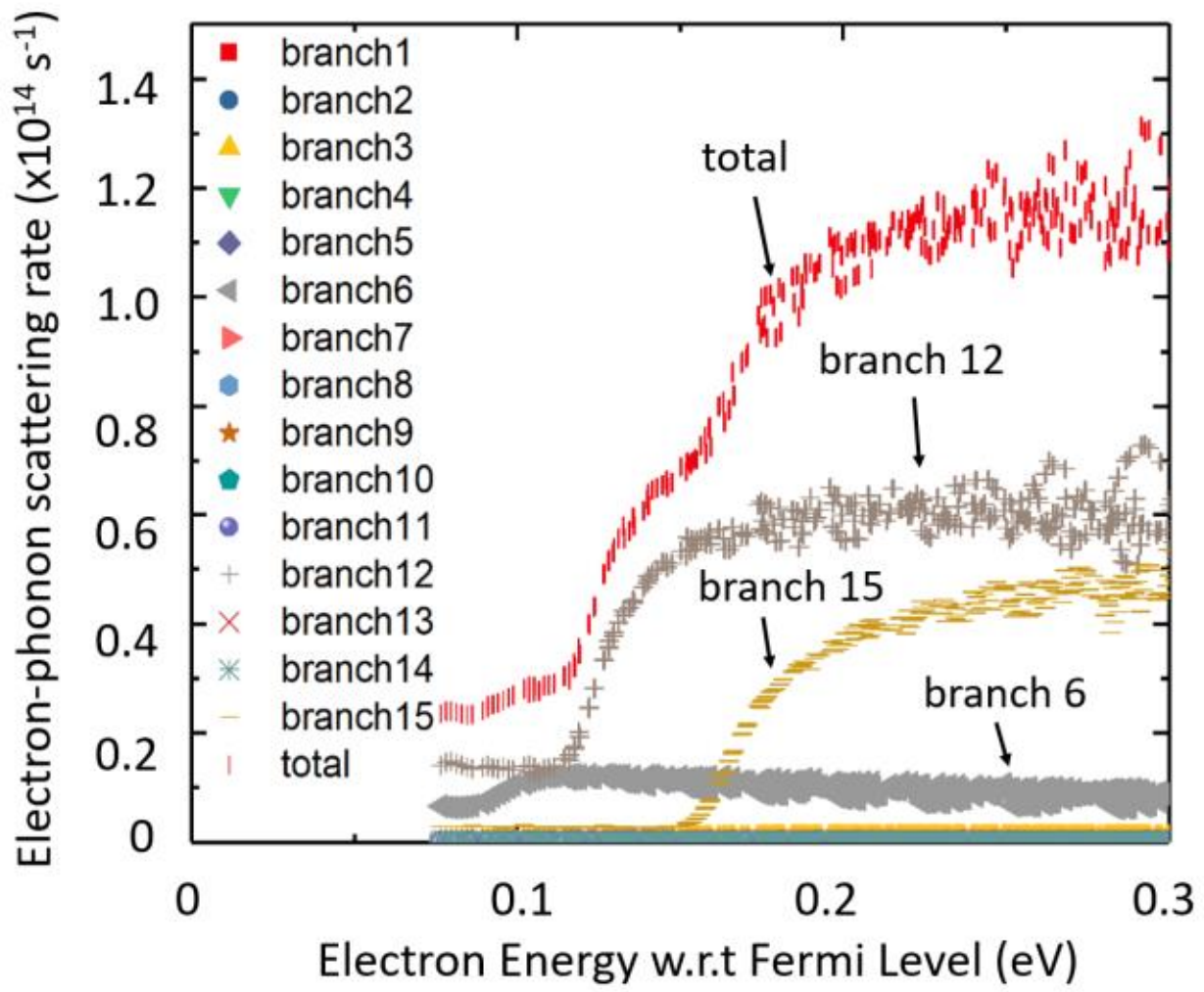

**FIG. 3.** Calculated electron-phonon scattering rates (in $s^{-1}$) in BSO as a function of electron energy with respect to the Fermi level (in eV). The LO phonons representing branches 6, 12 and 15 of the phonon dispersion spectra (also see supplementary material) play a dominant role in electron scattering with the total scattering rate increasing steeply beyond 120 meV.

Given the dominant optical phonon modes in BSO and strong LO-phonon induced carrier scattering shown above, the emission of LO-phonons by energetic carriers at the source is likely to act as the velocity saturation mechanism for BSO films. Such a model has been previously used to explain velocity saturation observed in GaN HEMTs, carbon nanotubes (CNTs) and layered materials.[25-28] To explain the model briefly, at high enough applied electric fields it becomes energetically favorable for carriers injected into the channel from the source to emit LO phonons and back-scatter back into the source region. This causes a net balance of carriers for forward injection leading to current and hence velocity saturation. The key feature of this model is the existence of a density-dependent saturation velocity as illustrated in Fig. 4 for the case of a 2-D electron gas with carrier density lower than the cross-over density ($n_0$) beyond which back-scattering into the bottom of the conduction band is Pauli blocked. At 0 K, the Fermi circle of radius $k_F$ indicates the range of occupied states for a given electron density. For smaller densities, we see from Fig.4 that the carriers can be accelerated to larger velocities (higher values of charge centroid, $k_0$) before the onset of optical phonon emission takes place ($k_{op}=k_F+k_0$) as compared to lower values for higher sheet carrier densities.

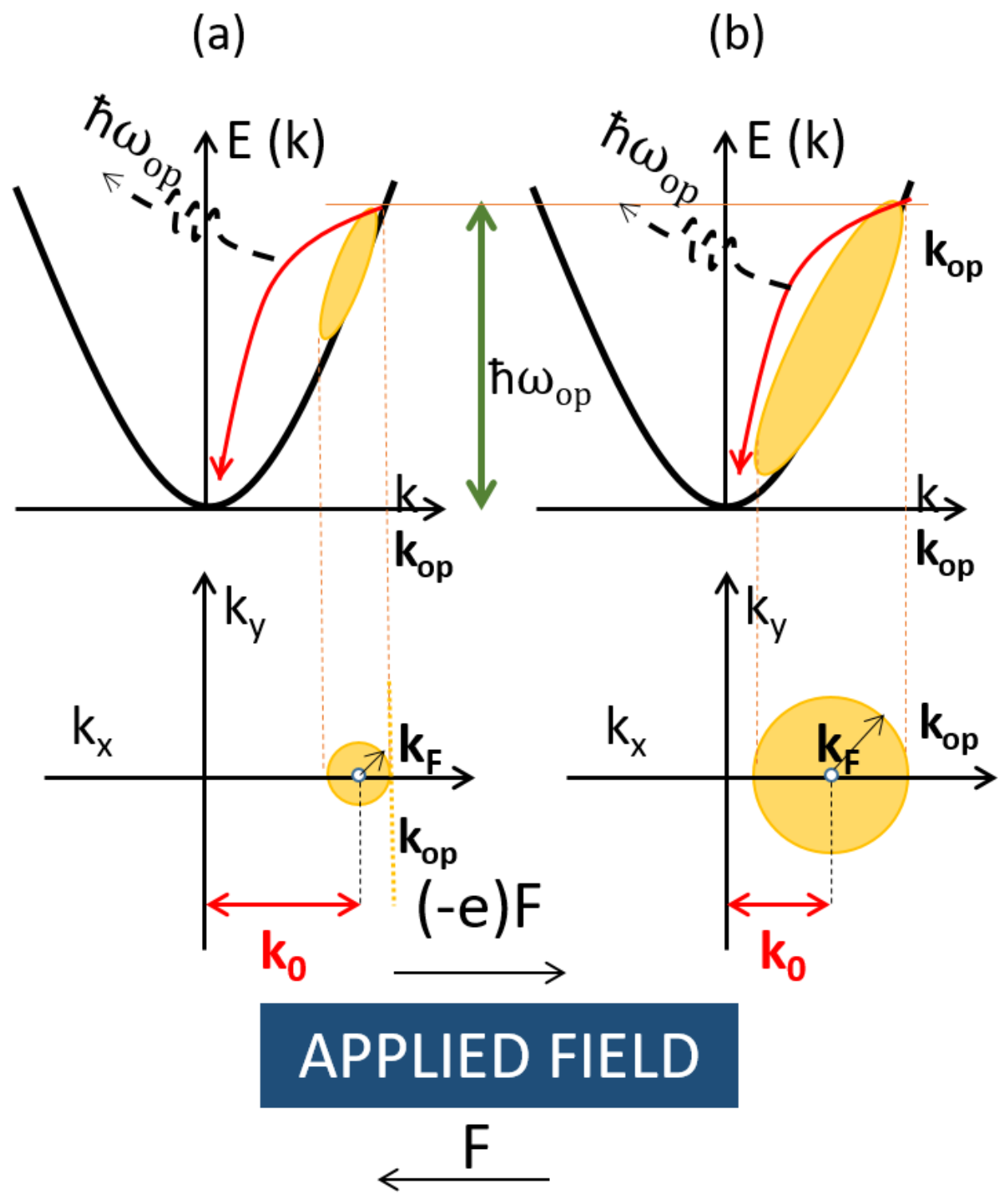

**FIG. 4.** Schematic showing a comparison of electron saturation velocity due to LO phonon emission for (a) smaller and (b) larger carrier densities in terms of energy band diagrams (top) and the corresponding k-space occupation of filled electrons (bottom) denoted by Fermi circles

($k_F$) at 0K for the 2D case. $k_0$ refers to the charge centroid under high applied fields when LO-phonon emission becomes favorable ($E(k)\geq\hbar\omega_{op}$).

The total current in the device can be estimated by summing over all occupied energy levels in k-space and is given by $J=qn_s(\hbar k_0/m^*)$ for electrons in a 2DEG, where $m^*$ is the effective mass ($0.2*m_0$ for BSO)[29] and the charge centroid $k_0$ is given by $m^*\omega_{op}/(2\hbar k_F)$ for $n_s>n_0$ and $k_0=k_{op}-k_F$ for $n_s\leq n_0$.[28] This gives rise to a square-root dependence between the current density and sheet carrier concentration for high densities ($n_0>n_s$) and since $J=qn_sv_{sat}$, $v_{sat}$ decreases as $n_s^{-1/2}$ with increasing channel charge density. The present study, however, involved samples of fairly-thick La-doped BSO films (16-21 nm) which more closely resembles a 3D electron gas rather than a 2D case even after accounting for the depletion widths in these layers. The current density for 3-D carriers can be estimated by summing the occupied states across a Fermi sphere of radius $k_F$ as,

$$\begin{aligned} J &= qn_{3D}t\frac{\hbar}{m^*}(3\pi^2(n_{0,3D}-n_{3D}))^{\frac{1}{3}}, n_{3D}\leq n_{0,3D} \\ &= \frac{qn_{3D}tE_{op}}{2\hbar(3\pi^2 n_{3D})^{1/3}}, n_{3D}>n_{0,3D} \end{aligned} \quad (1)$$

where $n_{0,3D}$ ($=k_{op}^3/3\pi^2=1.68\times10^{19}$ cm$^{-3}$) is the cross-over 3D density for which optical phonon emission to the bottom of the conduction band is not Pauli blocked and $E_{op}$ is the optical phonon energy (120 meV as discussed above). The saturation velocity extracted ($v_{sat}=J/qn_s$) from these equations for the 3D case therefore has a $n_{3D}^{-1/3}$ dependence at higher carrier densities in contrast to the $n_s^{-1/2}$ dependence for 2D confinement which also leads to higher values of $v_{sat}$ in comparison.

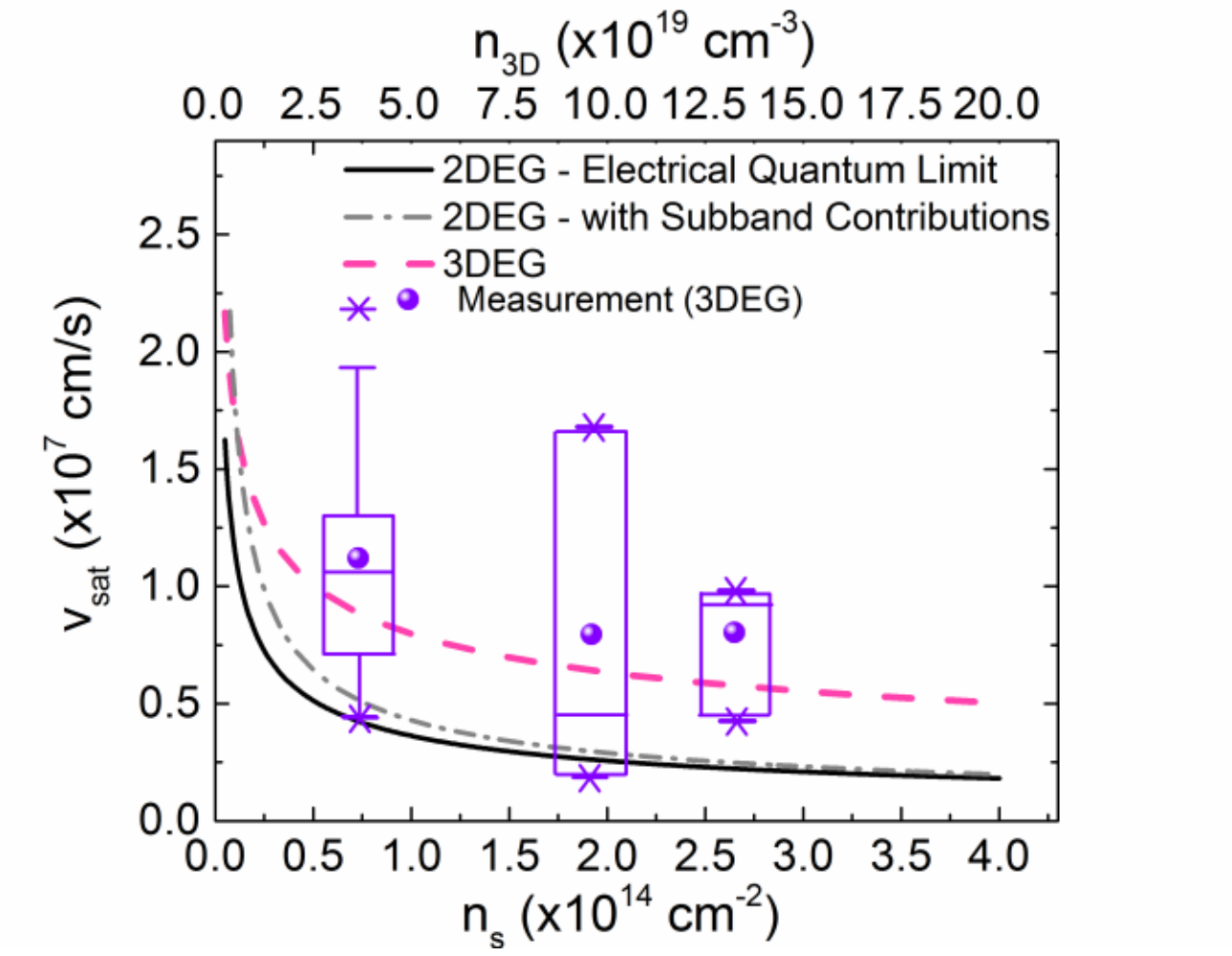

**FIG. 5.** LO phonon-limited saturation velocity of electrons for La-doped BSO 3DEG ($t_{BSO}$ =20 nm) and experimentally measured values represented as box plots for the La:BSO films of Table I shown for comparison. Also shown are the density-dependent $v_{sat}$ curves for δ-doped BSO 2DEGs in the electrical quantum limit (single sub-band) and multiple sub-band contributions ($t_{BSO}$ = 5 nm with 2 eV barriers on either side).

Fig. 5 shows the predicted saturation velocity for a range of equivalent sheet carrier concentration of the La:BSO 3DEGs along with experimentally measured values for comparison ($n_{3D}$ was estimated as $n_s/t_{La:BSO}$ for each case as the depletion widths in the La:BSO layer is <2.5 nm and ~5 nm respectively for the highest and lowest doping density films, and error bars were extracted from the measured device-to-device variation on each sample, see Fig. 2(a) and Fig. S2 in supplementary material). We see that the experimentally measured saturation velocity for a range of carrier concentrations shows reasonable agreement with the predictions of the 3-D optical phonon model (at 0 K), indicating that optical phonon emission using an effective LO phonon energy of 120 meV is the likely cause for velocity saturation in La:BSO thin films. It can also be seen from Fig. 5 that the measured saturation velocity decreases for higher carrier densities as predicted by the model. The use of modulation-doped barriers to $BaSnO_3$ channels, to spatially separate out the 2DEG in the high-mobility channel from the impurity atoms in the barrier, is another promising approach that has been mooted for device applications.[6] Hence we also estimate the density-dependence of saturation velocity for the 2-D confinement case corresponding to δ-doped BSO channels. Since the carrier concentrations in La:BSO are typically much higher (>$5\times10^{13}$ $cm^{-2}$) than those observed in other high charge density interfaces such as AlGaN/GaN, it is important to consider not just the occupancy of the first sub-band (the electrical quantum limit), but also of higher order sub-bands which contribute to current densities and hence saturation velocity. In order to account for filling multiple sub-bands at realistic doping densities, a δ-doped BSO channel of thickness 5 nm was considered with confining barriers of 2 eV on either side, analogous to a finite potential well where each bound state represents a sub-band. The total current density was then obtained by summing up the contributions from each sub-band and the net saturation velocity is also shown in Fig. 5 (see supplementary material for more details).

The saturation velocity for δ-doped BSO channels is estimated to drop from $1.6\times10^{7}$ cm/s to a more or less constant value of $2\times10^{6}$ cm/s for sheet carrier concentrations beyond $2\times10^{14}$ $cm^{-2}$. It is evident that even though the saturation velocity drops with increasing carrier densities, the decrease is not as steep (proportional to $n_s^{1/2}$) as the corresponding rise in charge density (~$n_s$ and hence the current drive) of La:BSO thin film transistors. Hence there is a definite motivation to have highly-doped BSO channels for high charge density switches with the modulation of such large charge densities being the primary challenge for device applications.

In conclusion, we have experimentally measured the saturation velocity of La:doped-$BaSnO_3$ thin films for a range of achievable doping densities. The experimental values show good agreement with the predictions of a LO-phonon emission model with an effective phonon energy of 120 meV, indicating the likelihood of optical phonon emission being the mechanism limiting velocity saturation in La:BSO thin films. Delta-doped BSO channels exhibit a density-dependent saturation velocity decreasing proportional to $n_s^{-1/2}$ from $1.6\times10^7$ cm/s to $2\times10^6$ cm/s at sheet carrier densities ranging from $10^{13}$ $cm^{-2}$ – $2\times10^{14}$ $cm^{-2}$, accounting for the contribution of multiple sub-bands necessary for such high carrier densities. The estimated and measured saturation velocities presented in this work would aid in informed device design for electronics based on thin film BSO channels.

**Supplementary Material**

Please see supplementary material for SEM images of representative constriction structures fabricated on each sample, the electric field vs current density plots the complete set of devices measured across all three samples used in this study along with tabulated data, electric field vs current density plots at 80 K and 300 K for the sample with $n_s = 7.23\times10^{13}$ $cm^{-2}$, the calculated phonon spectra of $BaSnO_3$ and details of the optical phonon model for the 2-D case with and without sub-band contributions.

The authors acknowledge funding from DARPA DREaM program (ONR N00014-18-1-2034, Program Manager Dr. Young-Kai Chen, monitored by Office of Naval Research, Program Manager Dr. Paul Maki), Office of Naval Research Grant N00014-18-1-2704 (Program Manager Dr. Brian Bennett), Semiconductor Research Corporation grant 2018-NC-2761-B and NSF ECCS- 1740119. T.W. and A.J. acknowledge support from National Science Foundation Faculty Early Career Development Program Grant No. DMR-1652994, and the Extreme Science and Engineering Discovery Environment (XSEDE) facility, National Science Foundation grant number ACI-1053575.

## REFERENCES:

[1] H. Y. Hwang, Y. Iwasa, M. Kawasaki, B. Keimer, N. Nagaosa and Y. Tokura, Nature Materials **11**, 103 (2012).

[2] S. Ramanathan, *Thin Film Metal-Oxides: Fundamentals and Applications in Electronics and Energy*, (Harvard University: Springer, 2010).

[3] H. J. Kim, U. Kim, H. M. Kim, T. H. Kim, H. S. Mun, B.-G. Jeon, K. T. Hong, W.-J. Lee, C. Ju, K. H. Kim and K. Char, Applied Physics Express **5** (6), 061102 (2012).

[4] H.-R. Liu, J.-H. Yang, H. J. Xiang, X. G. Gong and S.-H. Wei, Applied Physics Letters **102** (11), 112109 (2013).

[5] K. Krishnaswamy, B. Himmetoglu, Y. Kang, A. Janotti and C. G. Van de Walle, Physical Review B **95** (20), 205202 (2017).

[6] K. Krishnaswamy, L. Bjaalie, B. Himmetoglu, A. Janotti, L. Gordon and C. G. Van de Walle, Applied Physics Letters **108** (8), 083501 (2016).

[7] C. Park, U. Kim, C. J. Ju, J. S. Park, Y. M. Kim and K. Char, Applied Physics Letters **105**, 203503 (2014).

[8] J. Yue, A. Prakash, M. C. Robbins, S. J. Koester and B. Jalan, ACS Applied Materials and Interfaces **10**, 21061-21065 (2018).

[9] V. R. S. K. Chaganti, A. Prakash , J. Yue, B. Jalan and S. J. Koester, IEEE Electron Device Letters **39**(9), 1381-1384 (2018).

[10] K. Natori, Journal of Applied Physics **76** (8), 4879-4890 (1994).

[11] M. Lundstrom, IEEE Electron Device Letters **18** (7), 361-363 (1997).

[12] S. Raghavan, T. Schumann, H. Kim, J. Y. Zhang, T. A. Cain and S. Stemmer, APL Materials **4** (1), 016106 (2016).

[13] B. Jalan, R. Engel-Herbert, N. J. Wright, and S. Stemmer, J. Vac. Sci. Technol. A **27**, 461 (2009).

[14] S. Bajaj, O. F. Shoron, P. S. Park, S. Krishnamoorthy, F. Akyol, T.-H. Hung, S. Reza, E. M. Chumbes, J. Khurgin and S. Rajan, Applied Physics Letters **107** (15), 153504 (2015).

[15] A. Prakash, J. Dewey, H. Yun, J. S. Jeong, K. A. Mkhoyan and B. Jalan, Journal of Vacuum Science & Technology A **33**, 060608 (2015).

[16] H. Paik, Z. Chen, E. Lochocki, A. Seidner, A. Verma, N. Tanen, J. Park, M. Uchida, S. Shang, B.-C. Zhou, M. Br¨utzam, R. Uecker, Z.-K. Liu, D. Jena, K. M. Shen, D. A. Muller and D. G. Schlom, APL Materials **5**, 116107 (2017).

[17] T. Stanislavchuk, A. Sirenko, A. Litvinchuk, X. Luo and S.-W. Cheong, Journal of Applied Physics **112** (4), 044108 (2012).

[18] P. Hohenberg and W. Kohn, Physical Review **136**, B864 (1964).

[19] W. Kohn and L.J. Sham, Physical Review **140**, A1133 (1965).

[20] S. Baroni, P. Giannozzi, and A. Testa, Physical Review Letter **58**, 1861 (1987).

[21] P. Giannozzi, S. Baroni, N. Bonini, M. Calandra, R. Car, C. Cavazzoni, D. Ceresoli, G.L. Chiarotti, M. Cococcioni, I. Dabo, A. Dal Corso, S. de Gironcoli, S. Fabris, G. Fratesi, R. Gebauer, U. Gerstmann, C. Gougoussis, A. Kokalj, M. Lazzeri, L. Martin-Samos, N. Marzari, F. Mauri, R. Mazzarello, S. Paolini, A. Pasquarello, L. Paulatto, C. Sbraccia, S. Scandolo, G. Sclauzero, A.P. Seitsonen, A. Smogunov, P. Umari, and R.M. Wentzcovitch, Journal of Physics: Condensed Matter **21**, 395502 (2009).

[22] J.P. Perdew, K. Burke, and M. Ernzerhof, Physical Review Letter **77**, 3865 (1996).

[23] D.R. Hamann, Physical Review B **88**, 085117 (2013).

[24] S. Poncé, E.R. Margine, C. Verdi, and F. Giustino, Computer Physics Communications **209**, 116 (2016).
[25] D. Jena, Journal of Applied Physics **105** (12), 123701 (2009).
[26] X. Luo, Y. Lee, A. Konar, T. Fang, G. Xing, G. Snider and D. Jena, IEEE DRC Tech. Digest **67**, 29 (2008).
[27] H. Chandrasekar, K. L. Ganapathi, S. Bhattacharjee, N. Bhat and D. N. Nath, IEEE Transactions on Electron Devices **63** (2), 767-772 (2016).
[28] D. Jena and S. Rajan, arXiv preprint arXiv:1008.1154 (2010).
[29] S. James Allen, S. Raghavan, T. Schumann, K.-M. Law and S. Stemmer, Applied Physics Letters **108** (25), 252107 (2016).

# Supplementary Material:

# Velocity Saturation in La:doped $BaSnO_3$ Thin Films

Hareesh Chandrasekar,[1,a)] Junao Cheng,[1] Tianshi Wang,[3] Zhanbo Xia,[1] Nicholas G. Combs,[2] Christopher R. Freeze,[2] Patrick B. Marshall,[2] Joe McGlone,[1] Aaron Arehart,[1] Steven Ringel,[1,4] Anderson Janotti,[3] Susanne Stemmer,[2] Wu Lu[1] and Siddharth Rajan[1,4]

[1] Department of Electrical and Computer Engineering, The Ohio State University, Columbus, OH 43210, USA

[2] Materials Department, University of California, Santa Barbara, CA 93106, USA

[3] Department of Materials Science and Engineering, University of Delaware, Newark, DE 19716, USA

[4] Department of Material Science Engineering, The Ohio State University, Columbus, OH 43210, USA

a) Electronic Mail: chandrasekar.28@osu.edu

**Experimental Details :**

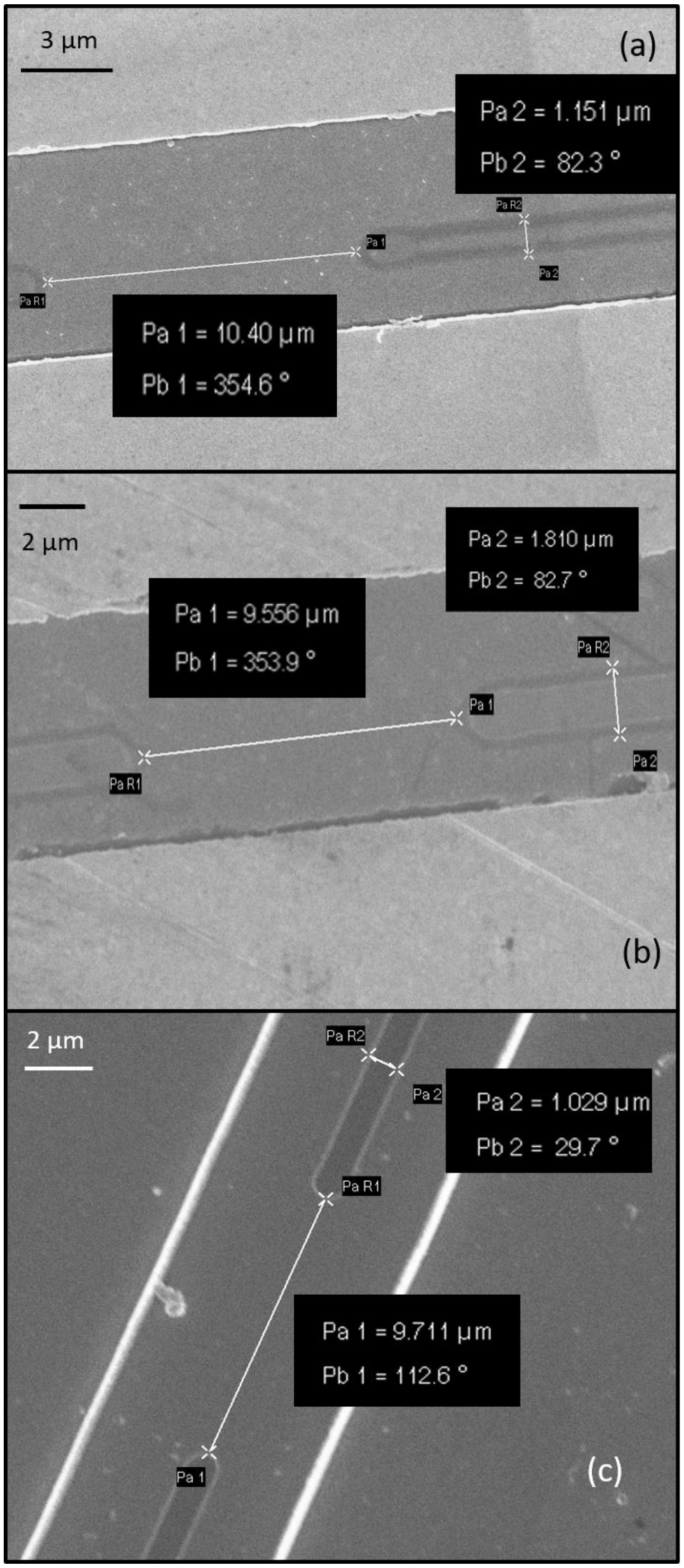

**FIG. S1.** SEM images of representative I-shaped constriction structures with indicated lengths and widths fabricated on La:doped $BaSnO_3$ films of sheet charge densities – (a) $2.65\times10^{14}$ $cm^{-2}$

($n_{3D}$ = 1.33×10$^{20}$ cm$^{-3}$), (b) 1.92×10$^{14}$ cm$^{-2}$ ($n_{3D}$ = 9.6×10$^{19}$ cm$^{-3}$) and (c) 7.23×10$^{13}$ cm$^{-2}$ ($n_{3D}$ = 3.65×10$^{19}$ cm$^{-3}$).

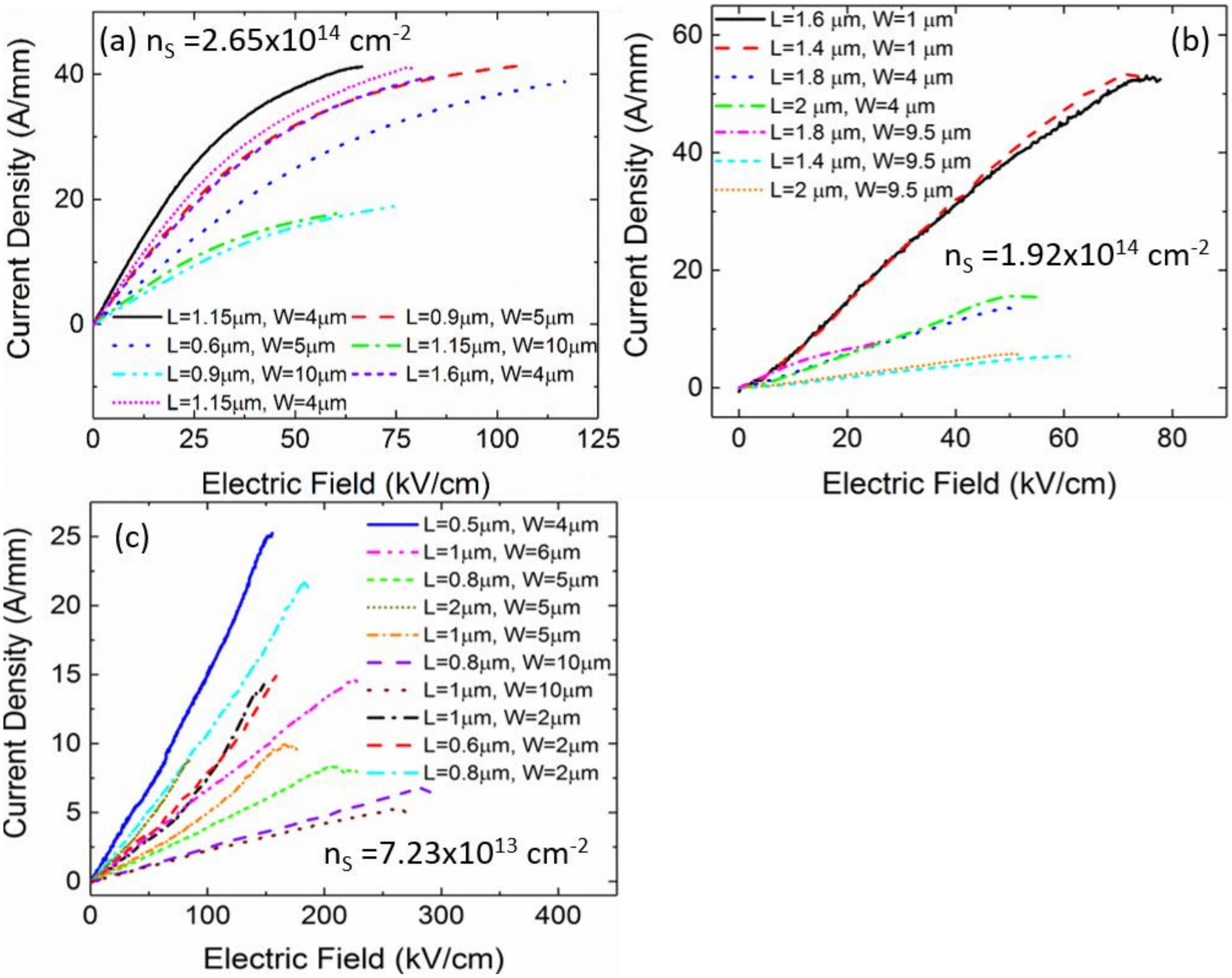

**FIG. S2.** Pulsed I-V measurements of I-shaped constriction structures of the indicated dimensions on 20 nm La:doped $BaSnO_3$ films of sheet charge density – (a) 2.65×10$^{14}$ cm$^{-2}$ ($n_{3D}$ = 1.33×10$^{20}$ cm$^{-3}$) – reproduction of Fig. 2 from main text, (b) 1.92×10$^{14}$ cm$^{-2}$ ($n_{3D}$ = 9.6×10$^{19}$ cm$^{-3}$) and (c) 7.23×10$^{13}$ cm$^{-2}$ ($n_{3D}$ = 3.65×10$^{19}$ cm$^{-3}$). Electric field (kV/cm) was estimated as applied voltage/constriction length and current density (A/mm) as measured current/constriction width.

The low-field drift mobility in Fig. S2(a) varies from 60.6-157.8 cm$^2$/Vs while the Hall mobility was measured to be 96 cm$^2$/Vs as indicated in Table I of main text. The Hall mobility provides an average value as the Hall test structures used were large (240 μm diameter circular van der Pauw structure) as compared to the highly scaled constriction structures used for saturation velocity measurement (at its largest L×W of 2 μm ×10 μm). Similarly for the other two samples in this study the drift mobilities were 164.1-24 $cm^2/Vs$ from Fig. S2(b) ($n_s$ = 1.92×$10^{14}$ $cm^{-2}$) and 122.8-18.6 $cm^2/Vs$ from Fig. S2(c) ($n_s$=7.23×$10^{13}$ $cm^{-2}$) while their Hall mobilities are respectively, 87 and 58 $cm^2/Vs$.

A list of all the test structures used for velocity measurement on each sample, the measured max current density $J_{max}$ and extracted $v_{sat}$ (from all devices of the three samples in Fig. S2) have also been tabulated below with their standard errors for easy reference. This data has been used to generate the box plot for measured saturation velocity in Fig. 5 of main text.

| $n_s$ = 2.65×$10^{14}$ $cm^{-2}$ , μ = 96 $cm^2/Vs$ (Fig. S1 (a)) | | |
|---|---|---|
| Device Dimensions | $J_{max}$ (A/mm) | $v_{sat}$ (cm/s) |
| L=1.15 μm, W=4 μm | 41.22 | 9.72×$10^6$ |
| L=1.15 μm, W=4 μm | 41.55 | 9.80×$10^6$ |
| L=0.6 μm, W=5 μm | 38.81 | 9.15×$10^6$ |
| L=1.15 μm, W=10 μm | 17.72 | 4.18×$10^6$ |
| L=0.9 μm, W=10 μm | 18.88 | 4.45×$10^6$ |
| L=1.6 μm, W=4 μm | 39.54 | 9.33×$10^6$ |
| L=0.9 μm, W=5 μm | 40.92 | 9.65×$10^6$ |
| Average | | 8.04×$10^6$ |
| Standard Deviation | | 2.56×$10^6$ |
| Standard Error | | 9.66×$10^5$ |
| $n_s$ = 1.92×$10^{14}$ $cm^{-2}$ , μ = 87 $cm^2/Vs$ (Fig. S2 (b)) | | |
| Device Dimensions | $J_{max}$ (A/mm) | $v_{sat}$ (cm/s) |
| L=1.6 μm, W=1 μm | 52.48 | 1.71×$10^7$ |
| L=1.4 μm, W=1 μm | 53.12 | $1.73\times10^7$ |
| L=1.8 μm, W=4 μm | 13.8 | $4.49\times10^6$ |
| L=2 μm, W=4 μm | 15.9 | $5.18\times10^6$ |
| L=2 μm, W=9.5 μm | 5.8 | $1.89\times10^6$ |
| L=1.4 μm, W=9.5 μm | 5.35 | $1.74\times10^6$ |
| L=1.8 μm, W=9.5 μm | 7.81 | $2.54\times10^6$ |
| Average | | $7.17\times10^6$ |
| Standard Deviation | | $6.96\times10^6$ |
| Standard Error | | $2.98\times10^6$ |
| $n_s = 7.23\times10^{13}$ cm$^{-2}$ , μ = 58 cm$^2$/Vs (Fig. S2 (c)) | | |
| Device Dimensions | $J_{max}$ (A/mm) | $v_{sat}$ (cm/s) |
| L=1 μm, W=6 μm | 14.71 | $1.27\times10^7$ |
| L=0.8 μm, W=5 μm | 8.33 | $7.2\times10^6$ |
| L=2 μm, W=5 μm | 8.76 | $7.57\times10^6$ |
| L=1 μm, W=5 μm | 9.95 | $8.6\times10^6$ |
| L=0.8 μm, W=10 μm | 6.78 | $5.86\times10^6$ |
| L=1 μm, W=10 μm | 5.26 | $4.55\times10^6$ |
| L=1 μm, W=2 μm | 14.33 | $1.24\times10^7$ |
| L=0.6 μm, W=2 μm | 14.9 | $1.29\times10^7$ |
| L=0.5 μm, W=4 μm | 25.27 | $2.18\times10^7$ |
| L=0.8 μm, W=2 μm | 21.8 | $1.88\times10^{7}$ |
| Average | | $1.12\times10^{7}$ |
| Standard Deviation | | $5.65\times10^{6}$ |
| Standard Error | | $1.79\times10^{6}$ |

**Table S1.** Maximum current density ($J_{max}$ in A/mm, from Fig. S2 above) and extracted saturation velocity (in cm/s) along with the average, standard deviation and standard errors, and measured device dimensions (from SEM) for all devices across the three samples used in this study. Full current density vs electric field curves are plotted in Fig. S2.

A comparison of the measured current density vs electric field curves at 300 K and 80 K for three devices on the lowest doped sample used in this study ($n_s$ =$7.23\times10^{13}$ cm$^{-2}$) is shown below in Fig. S3 indicating a maximum change of 6% in $J_{max}$ (and hence $v_{sat}$) in these films indicating that the saturation is velocity-dominated and not mobility-limited.

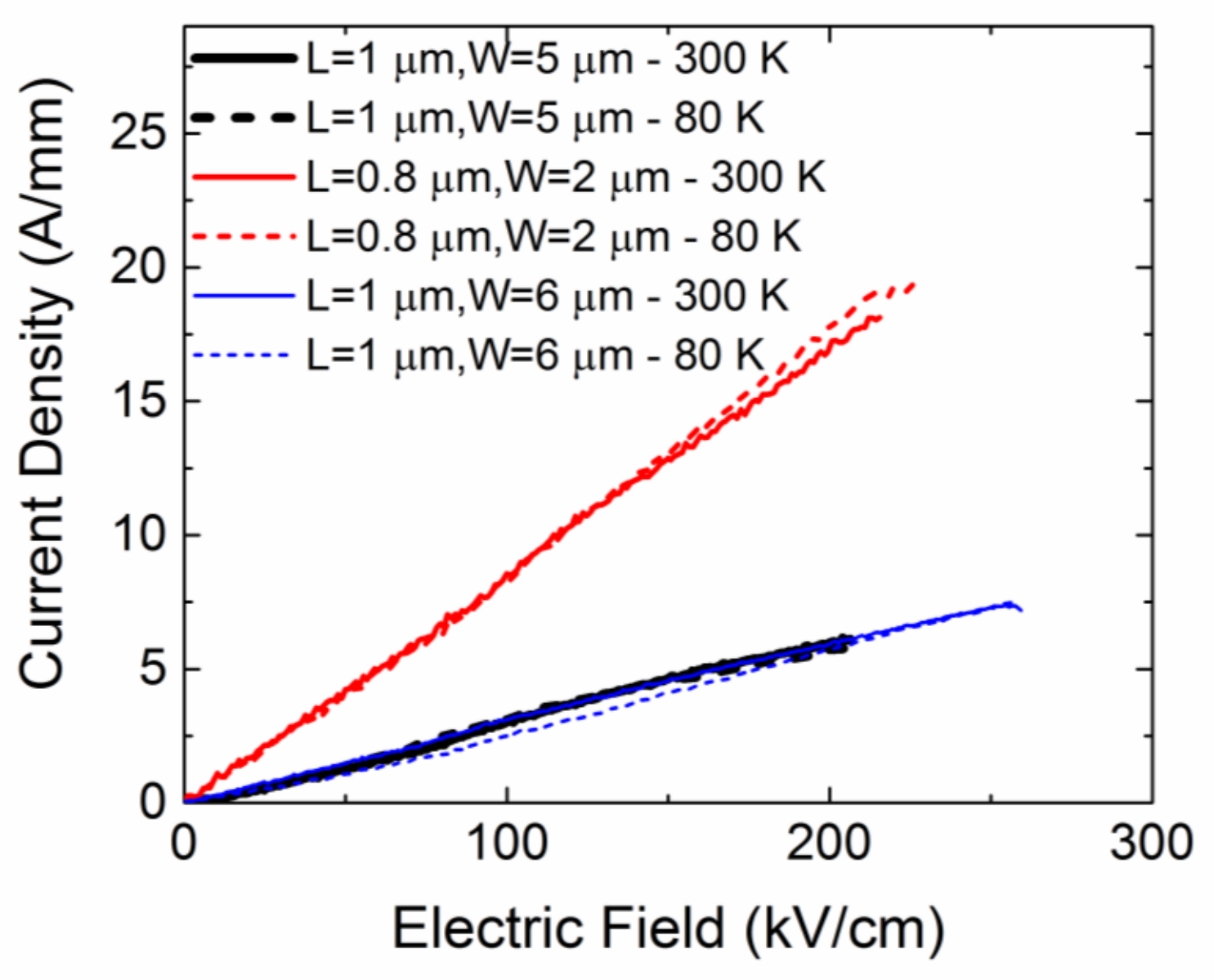

**FIG. S3.** Current Density vs Electric Field plots for the lowest ($n_s$ =$7.23\times10^{13}$ cm$^{-2}$) doping density sample used in this study for indicated device dimensions measured at 300 K and 80 K.

## Calculated Phonon Spectra of BSO:

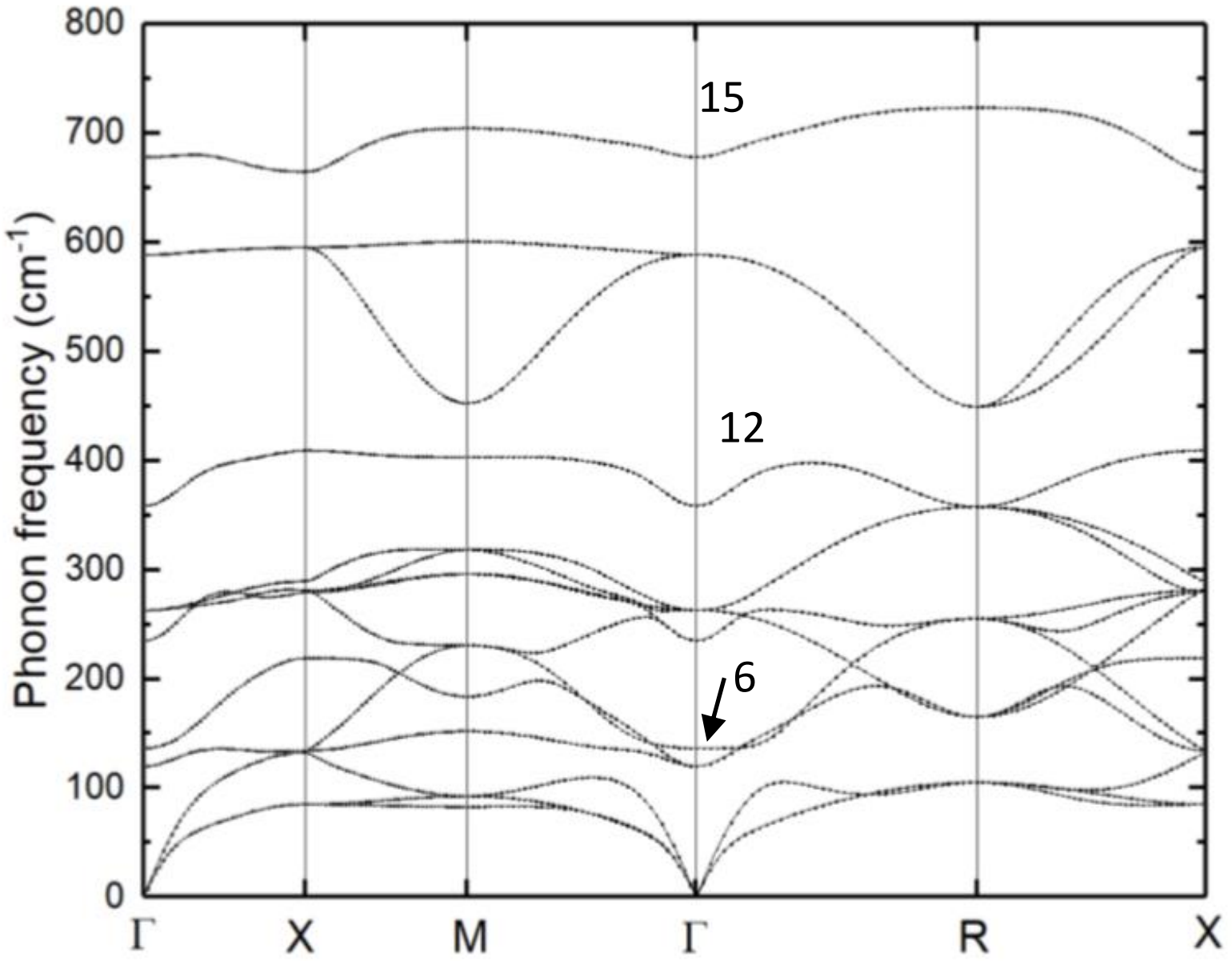

**FIG. S4.** Calculated phonon spectrum of $BaSnO_3$ based on density functional perturbation theory (DFPT)[2] as implemented in the Quantum ESPRESSO code[3].

## Details of the optical phonon model:

<u>2D-case:</u>

Current density can be obtained by summing across occupied states within a Fermi circle of radius $k_f$ (=$\sqrt{(2\pi n_s)}$ at 0K) which for the case of a parabolic band reduces to,

$$J = \frac{2q}{(2\pi)^2} \int d^2k \left(\frac{\hbar k_0}{m^*}\right) f(k) \qquad \text{(S1)}$$

, where $k_0$ is the centroid of the Fermi circle which shifts under applied bias and other terms have their usual meanings.

For carrier densities beyond cross-over 2DEG density beyond $n_0$ (=$k_{op}^2/8\pi=2.5\text{x}10^{12}$ $cm^{-2}$) optical phonon emission becomes favorable and current densities have been shown to be:

$$J = qn_s \frac{\hbar\left(k_{op} - \sqrt{2\pi n_s}\right)}{m^*} , n_s \leq n_0 \qquad \text{(S2)}$$

$$= q\,\omega_{op}\sqrt{\frac{n_s}{8\pi}}\,, n_s > n_0$$

2D-case with sub-bands:

The occupancy of each sub-band was estimated as $n_{s,\text{subband } i} = m^*/(\pi\hbar^2)\,(E_F - E_i)$ where $E_i$ refers to the bound state energy of the $i^{th}$ sub-band and the current density due to each sub-band was then estimated as above and summed to obtain the total current density.

δ-doped $BaSnO_3$ layer of thickness 5 nm and having 2 eV barriers was treated analogous to a finite potential well problem as shown in the band-diagram below.

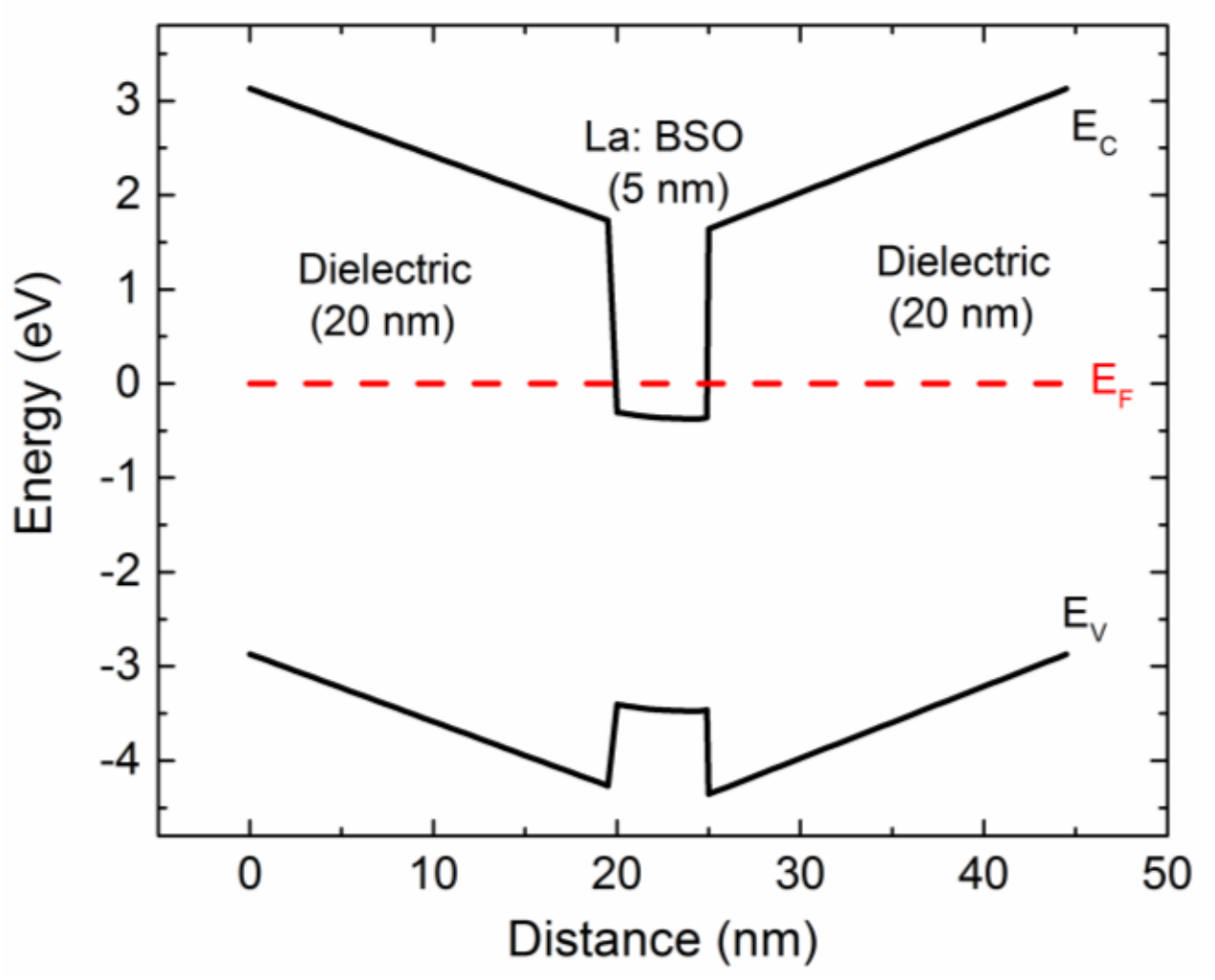

**FIG. S5.** Representative band diagram for an La:doped BSO channel ($10^{20}$ $cm^{-3}$ doping density) of 5 nm thickness sandwiches between two dielectric layers (ε=10,thickness=20 nm) having a conduction band offset of 2 eV with reference to the channel material.

The bound state solutions were obtained along the lines indicated by Chiani[1] (with energies of $E_1$=0.067 eV, $E_2$=0.133 eV, $E_3$=0.199 eV, $E_4$=0.265 eV, $E_5$=0.328 eV, $E_6$=0.412 eV) and were used in the current density expressions.

3D case:

Similar to the 2D case, the Fermi surface is a sphere at 0K for the case of a 3DEG in the thick La:doped BSO films measured here. The cross-over density for which optical phonon emission becomes favorable is then given by $n_0$ ($=k_{op}^3/3\pi^2$) and the current density itself can be estimated as,

$$J = \frac{2q}{(2\pi)^3}\int d^3k \left(\frac{\hbar k_0}{m^*}\right) f(k) \tag{S3}$$

, which then reduces to the equations given in the main text:

$$\begin{aligned} J &= qn_{3D}t\frac{\hbar}{m^*}(3\pi^2(n_{0,3D}-n_{3D}))^{\frac{1}{3}}, n_{3D} \le n_{0,3D} \\ &= \frac{qn_{3D}tE_{op}}{2\hbar(3\pi^2 n_{3D})^{1/3}}, n_{3D} > n_{0,3D} \end{aligned} \tag{S4}$$

**References:**

[1] *Chiani, M. (2016). "A chart for the energy levels of the square quantum well". arXiv:1610.04468*

[2] S. Baroni, P. Giannozzi, and A. Testa, Physical Review Letter **58**, 1861 (1987).

[3] P. Giannozzi, S. Baroni, N. Bonini, M. Calandra, R. Car, C. Cavazzoni, D. Ceresoli, G.L. Chiarotti, M. Cococcioni, I. Dabo, A. Dal Corso, S. de Gironcoli, S. Fabris, G. Fratesi, R. Gebauer, U. Gerstmann, C. Gougoussis, A. Kokalj, M. Lazzeri, L. Martin-Samos, N. Marzari, F. Mauri, R. Mazzarello, S. Paolini, A. Pasquarello, L. Paulatto, C. Sbraccia, S. Scandolo, G. Sclauzero, A.P. Seitsonen, A. Smogunov, P. Umari, and R.M. Wentzcovitch, Journal of Physics: Condensed Matter **21**, 395502 (2009).